\documentstyle[12pt]{article}
\def\dir{}

\def\figcond{1}

\def\figsty{0}

\def\authname{}

\newcount\figsubcountss \figsubcountss=0
\ifnum\figcond>0 
\input epsf 
  \ifnum\figsty>0 
  \fi
\fi
\def\figinsert#1#2#3#4{
\ifnum\figcond>0 
  \ifnum\figsty>0 
    \ifnum\figsubcountss=0
      \immediate\write9{\noexpand\catcode`\noexpand\@=11}
      \immediate\write9{\noexpand\newpage }	
      \immediate\write9{\noexpand\pagestyle {empty}}
      \immediate\write9{\noexpand\section* {Figure Captions}}
      \immediate\write9{\noexpand\begin {enumerate}}
      \immediate\write9{\noexpand\renewcommand {\noexpand\theenumi}{}}
      \immediate\write9{\noexpand\begin {enumerate}}
      \immediate\write9{\noexpand\renewcommand 
        {\noexpand\theenumii}{\noexpand\arabic {enumii}}}
      \immediate\write9{\noexpand\renewcommand {\noexpand\labelenumii}
        {Fig. \noexpand\arabic {enumii}:}}
      \immediate\write10{\noexpand\newpage}
      \ifnum\figsty>1
        \immediate\write10{\noexpand\pagestyle {headings}}
        \immediate\write10{\noexpand\setcounter {page}{1}}
        \immediate\write10{\noexpand\renewcommand {\noexpand\thepage}
            {Figure \noexpand\arabic{page} -- \noexpand\authname}}
      \else
        \immediate\write10{\noexpand\pagestyle {empty}}
      \fi
    \fi
    \global\advance\figsubcountss by 1
    \immediate\write10{\noexpand\begin {figure}[p]}
    \immediate\write10{\noexpand\begin {center}}
    \immediate\write10{\noexpand\ }
    \immediate\write10{\noexpand\epsfbox {\dir #1}}
    \immediate\write10{\noexpand\ \noexpand\\}
    \ifnum\figsty<2
      \immediate\write10{\noexpand\vspace {1cm}}
      \immediate\write10{Figure \noexpand\ref {#3}}
    \fi
    \immediate\write10{\noexpand\end {center}}
    \immediate\write10{\noexpand\end {figure}}
    \immediate\write10{\noexpand\newpage}
   \let\save=\ref \let\ref=0 \let\savec=\cite \let\cite=0
    \immediate\write9{\noexpand\item #2}
   \let\ref=\save \let\cite=\savec
    \immediate\write9{\noexpand\label {#3}}
    \begin {figure}[htbp]
    \begin{center}
    \fbox{Fig. \ref{#3}}
    \end{center}
    \end {figure}
  \else
    \begin{figure}[#4]
    \begin{center}
    \ \epsfbox{\dir #1}
    \caption []{#2 \label {#3}}
    \end{center}
    \end{figure}
  \fi
\else
  \begin {figure}[htbp]
  \begin{center}
  \fbox{Fig. \ref{#3}}
  \caption []{#2 \label {#3}}
  \end{center}
  \end {figure}
\fi
}
\def\Closeout#1{%
   \immediate\closeout#1}
\def\figepsfout{
\Closeout10
  \immediate\write9{\noexpand\end {enumerate}}
  \immediate\write9{\noexpand\end {enumerate}}
  \immediate\write9{\noexpand\catcode`\noexpand\@=12}
\Closeout9
\input \jobname.cap
\input \jobname.fis}
\catcode`\@=11
\newbox\tempboxa
\newdimen\captionboxsubcount 
\def\capsize#1{\captionboxsubcount=#1pt}
\newdimen\captionboxsub
\captionboxsub=\hsize \advance\captionboxsub by -\captionboxsubcount
\advance\captionboxsub by -\captionboxsubcount
\long\def\@makecaption#1#2{
 \setbox\@tempboxa\hbox{#1: #2}
 \ifdim \wd\@tempboxa >\captionboxsub 
\rightskip=\captionboxsubcount \leftskip=\captionboxsubcount #1: #2 
\else \hbox to\hsize{\hfil\box\@tempboxa\hfil} 
 \fi}
\def\enddocument{
\ifnum\figcond>0
  \ifnum \figsty>0 \figepsfout
\fi\fi
\@checkend{document}\clearpage\begingroup  
\if@filesw \immediate\closeout\@mainaux 
\def\global\@namedef##1##2{}\def\newlabel{\@testdef r}%
\def\bibcite{\@testdef b}\@tempswafalse \makeatletter\input \jobname.aux
\if@tempswa \@warning{Label(s) may have changed.  Rerun to get
cross-references right}\fi\fi\endgroup\deadcycles\z@\@@end}
\catcode`\@=12
\capsize{30}

\newcommand{\ba}{\begin{array}}  
\newcommand{\ea}{\end{array}}  
\newcommand{\bea}{\begin{eqnarray}}  
\newcommand{\eea}{\end{eqnarray}}  
\newcommand{\be}{\begin{equation}}  
\newcommand{\ee}{\end{equation}}  
\newcommand{\gapproxeq}{\lower .7ex\hbox{$\;\stackrel{\textstyle
>}{\sim}\;$}}  
\newcommand{\lapproxeq}{\lower .7ex\hbox{$\;\stackrel{\textstyle
<}{\sim}\;$}}  

\def\1N{\displaystyle{1/N_c}}

\begin{document}  


\setcounter{page}{1}

\begin{flushright}
\begin{tabular}{l}
SU-4240-650\\
\end{tabular}
\end{flushright}
\bigskip
\begin{center}
\LARGE
\bf 
Comment on ``Confirmation of the Sigma Meson'' 
\end{center}
\begin{center}
\large 
{\it Masayasu Harada $^{a}$
\footnote{{\it Electronic address:}~mharada@npac.syr.edu}}
~~~~~~~~~~{\it Francesco Sannino $^{a,b}$
\footnote{{\it Electronic address:}~sannino@npac.syr.edu,}} 
\end{center}
\begin{center}
\large {\it 
Joseph Schechter $^{a}$
\footnote{{\it Electronic address:}~schechter@suhep.phy.syr.edu}}
\end{center}
\begin{center}
\smallskip
$^a$~{Department of Physics, Syracuse University, Syracuse,
New York,
13244-1130.}
\end{center}
\begin{center}
\smallskip
$^b$~{Di\-par\-ti\-mento di Scienze Fi\-si\-che 
\& Istituto Nazionale di Fisica Nucleare, Mo\-stra d'Ol\-tre\-mare
Pad.\-19, 80125
Na\-po\-li, Ita\-lia.}
\end{center}
\begin{flushleft}
\end{flushleft}

In Ref.~[\ref{Tornqvist-Roos}], T\"ornqvist and Roos presented a model
of $\pi\pi$ scattering which supports the existence of the old
$\sigma$ meson at a mass of $397$ MeV and width $590$ MeV. While this
model is constructed to satisfy unitarity, it does not explicitly take
crossing symmetry into account. In particular, one may question
[\ref{Isgur-Speth}] the validity of neglecting the crossed-channel
$\rho$ meson exchange contributions, which are generally considered to
be important. It is actually very complicated, as noted by the authors
themselves, to examine this question in their model. Here we
investigate this issue in the framework of a recently proposed
[\ref{Harada-Sannino-Schechter}] simple model for $\pi\pi$
scattering. We find that the consistent neglect of the $\rho$ exchange
does not destroy the existence of the $\sigma$ meson but modifies its
parameters so that they get close to the results of
Ref.~[\ref{Tornqvist-Roos}].   

The simple model in question may be most conservatively regarded as an 
approximate parameterization of the relativistic $\pi\pi$ amplitude 
which satisfies both crossing symmetry and unitarity up to $1.2$ GeV. 
It is based on a chiral symmetric Lagrangian. The amplitude 
is constructed by adding together four components: 
(1) the {\it current algebra} contact term, 
(2) the $\rho$ exchange diagrams, 
(3) a $\sigma$ piece, 
(4) the $f_0(980)$ with an associated 
{\it Ramsauer-Townsend} mechanism. 
The only three parameters which must be fit to experiment appear 
in a {\it regularized} description of the 
real part of the pole term which is proportional to 
\begin{displaymath}
{\rm Re}\left[ \frac{M_{\sigma} G}
{M^2_{\sigma} - s - i M_{\sigma} G^{\prime}} \right] 
\ . 
\end{displaymath}
Note that, since 
$G \ne G^{\prime}$, this is not identical to a Breit-Wigner 
form for this very broad object. 
Figures 1 and 2 of Ref.~[\ref{Harada-Sannino-Schechter}] 
show that the $\sigma$ is absolutely essential to preserve unitarity.

In [\ref{Harada-Sannino-Schechter}] a best fit 
to the real part of the $I=J=0$ 
partial amplitude, $R^0_0$ was found for a mass 
$M_{\sigma}=559$ MeV, a width $G^{\prime}=370$ MeV 
and $G/G^{\prime}=0.29$. It is an easy matter 
to neglect the $\rho$ meson contributions 
(including the associated contact term needed for chiral symmetry) 
and make a new fit. The resulting $R^0_0$ in comparison 
with the experimental data is shown in 
Fig.~1 and is about as good as the previous fit including 
the $\rho$ meson. (Of course, the $\rho$ meson is definitely present 
in nature.) The new fitted parameters are the mass 
$M_{\sigma}=378$ MeV, the width $G^{\prime}=836$ MeV and 
$G/G^{\prime}=0.08$. The new mass and width are close 
to the values found in Ref.~[\ref{Tornqvist-Roos}]. 
We therefore would expect that including the $\rho$ exchange 
in their framework would raise their mass by roughly 
$150$ MeV and lower their width prediction. 
This behavior can be easily understood in a qualitative sense, 
since the addition of the $\rho$ raises the energy at 
which the unitarity bound is violated 
(see Fig.~1 of Ref.~[\ref{Harada-Sannino-Schechter}]). 
Of course, in Ref.~[\ref{Harada-Sannino-Schechter}], 
the question of whether the $\sigma$ and $f_0(980)$ 
are $q\bar{q}$, $q^2 \bar{q}^2$ states or 
some superposition is not directly addressed.

\vskip 1 cm

\begin{enumerate}

\renewcommand{\labelenumi}{[\arabic{enumi}]}

\item N.A.~T\"ornqvist and M.~Roos, Phys. Rev. Lett. {\bf 76}, 
1575 (1996).
\label{Tornqvist-Roos} 

\item {N.~Isgur and J.~Speth, Phys. Rev. Lett. 
{\bf 77}, 2332 (1996); 
N.A.~T\"ornqvist and M.~Roos, 
{\it ibid}, 2333 and references therein.}
\label{Isgur-Speth} 

\item {M.~Harada, F.~Sannino and J.~Schechter, 
Phys. Rev. D {\bf 54}, 1991 (1996).}
\label{Harada-Sannino-Schechter}

\item {E.A. Alekseeva et al., Soc. Phys. JETP {\bf 55},
591 (1982).}
\label{lowenergy}

\item {Grayer et al., Nucl. Phys. {\bf B75}, 189 (1974)}
\label{highenergy}

\end{enumerate}

\figinsert{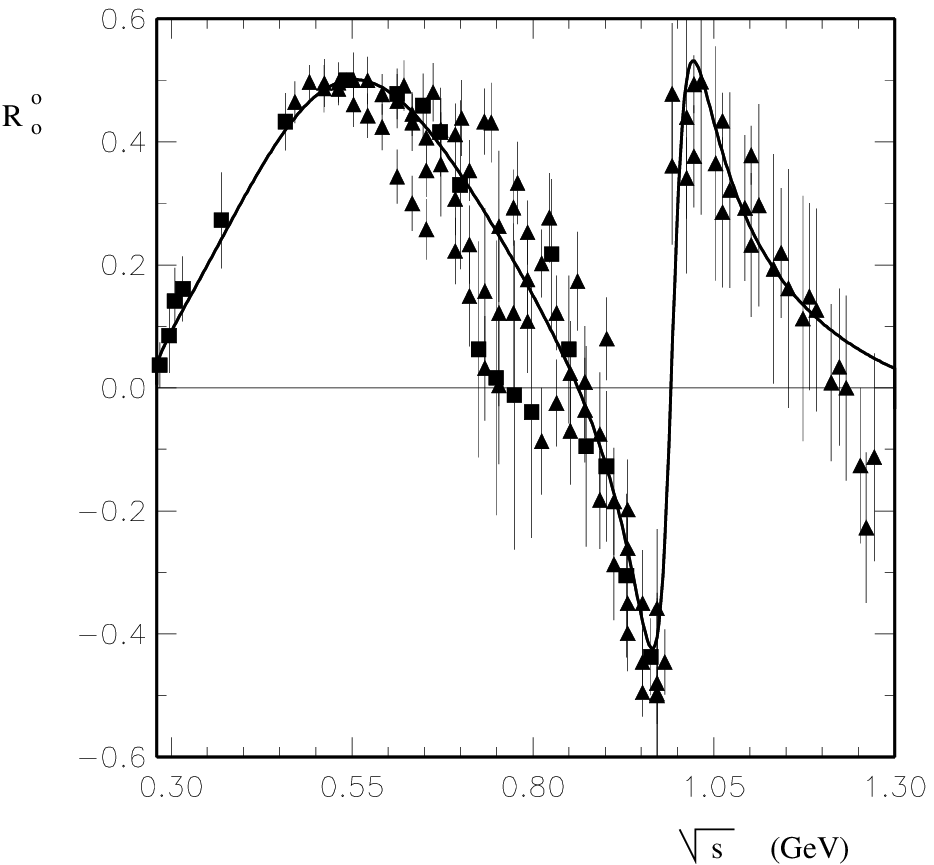}{The solid line is the {\it current algebra}
$+~\sigma +~f_0(980)$ result for $R^0_0$. The experimental points are
extracted from the measured phase shifts by neglecting the small 
inelasticity effects. ($\Box$) are extracted from
the data of Ref.~4 while ($\triangle$) are extracted from the data of Ref.~5}
{figuracomment}{htpb}

\end{document}